\documentclass[twocolumn]{svjour3}          
\smartqed                                   

\usepackage{amsmath}
\usepackage{amssymb}
\usepackage{graphicx}

\newcommand{\fd}{\dot f}
\newcommand{\fdd}{\ddot f}

\journalname{Eur. Phys. J. C}

\begin{document}

\title{Constraining the linear equation of state during shear-free radiative collapse}
\titlerunning{Constraining the linear equation of state}

\author{Megandhren Govender}
\authorrunning{M. Govender}

\institute{M. Govender \at
  Department of Mathematics, Faculty of Applied Sciences, Durban University of Technology, Durban 4000, South Africa \\
  \email{megandhreng@dut.ac.za}}

\date{Received: date / Accepted: date}

\maketitle

\begin{abstract}
We investigate the constraints on the equation of state (EoS) parameter for any shear-free, spherically symmetric
radiating star with separable metric potentials. Imposing the linear EoS, $p_r=\alpha\rho-\eta$, throughout the stellar interior, together with the Santos
junction condition, forces $\alpha=-1/3$, independently of the choice of initial static configuration. For constant $\eta$ this requires $\eta$ to vanish. On the other hand, if $\eta$ is allowed to evolve with time, the ensuing model is incompatible with the generic
solution of the junction equation. Models that quote results for several values of $\alpha$
while imposing the linear EoS on the dynamical fluid are therefore internally
inconsistent. A consistent reading restricts the equation of state to the initial static
configuration, where $\alpha$ is a free parameter fixing $g_{tt}$ for the initial static progenitor.
\end{abstract}

\section{Introduction}
Since the presentation of the Santos junction conditions for nonadiabatic gravitational collapse, the evolution of a radiating star matched to a Vaidya exterior has become a framework for studying the interplay of gravity, heat flow and thermodynamics \cite{opp,herrera,vaidya,santos}. Dissipative gravitational collapse has given researchers rich insights into the evolution of temperature, horizon formation, stability and, more recently, complexity in radiating stars. Early models employed shear-free matter distributions with isotropic pressure and heat flow.

A particularly fruitful strategy was utilized by de Oliveira et al. \cite{dok85}, in which the star departs from a static configuration in the infinite past and the metric potentials bifurcates into a static part and a time-dependent part. This scheme was developed further by de Oliveira et al. \cite{dks88}. It can be shown through the Raychaudhuri equation, shear-free collapse is the slowest possible mode of collapse \cite{bonnor}, so that such models are natural descriptions of the early, quasi-static stages of dissipative collapse. An important special case is acceleration-free (geodesic) collapse, studied by Kolassis et al.~\cite{kst88}, in which $g_{tt}$ is constant and the Friedmann dust solution is regained when the heat flux vanishes. Later, Ivanov \cite{ivanov} showed that the junction condition for geodesic, anisotropic collapse with shear and heat radiation reduces to a Riccati equation and obtained all of its solutions in terms of a generating function.

The assumption of shear-free motion has since been relaxed in several directions leading to more realistic models of dissipative collapse. Chan et al. \cite{chs93} studied the dynamical stability of radiating anisotropic collapse, and Chan \cite{chan97,chan03} developed collapse models with shear, an approach later refined by Pinheiro and Chan \cite{pc10}. Exact shearing solutions with geodesic trajectories, together with their causal thermal evolution, were presented by Naidu, Govender and Govinder \cite{naidu06}, while Govender et al. \cite{grm14} showed that shear modifies the collapse rate and enhances the core temperature. Ivanov \cite{ivanov16} studied shearing collapse by utilizing the horizon function, which reduces the junction condition to a more tractable form and allows its solutions to be related to physical features of the star, and later showed that shear and bulk viscosity, null radiation and charge can be absorbed into an effective anisotropic model with heat flow \cite{ivanov19}. More recently, Majozi et al. \cite{mmg26} constructed models for radiating stars with shear. A second family of simple models is that of Euclidean stars, in which the areal radius equals the proper radial distance throughout the fluid \cite{hs10,gg12,ggg10,abebe14,hdo25}. For these stars, dissipation by heat flow requires a nonvanishing four-acceleration, so dissipative Euclidean stars cannot be geodesic.

The question of how complicated the stellar interior needs to be has recently been given a more precise formulation through the notion of complexity. Herrera \cite{h18} proposed that the simplest fluid distribution is homogeneous in energy density with isotropic pressure, and introduced a complexity factor $Y_{TF}$ which emerges from the orthogonal splitting of the Riemann tensor. For time-dependent dissipative systems, Herrera, Di Prisco and Ospino \cite{hdo18} extended this definition and proposed the homologous condition as the simplest pattern of evolution; a vanishing complexity factor combined with homologous evolution forces the fluid to be geodesic. The less restrictive quasi-homologous condition was subsequently considered in Ref.~\cite{hdo20}, and the connection between complexity, homologous evolution and Euclidean stars is explored in Ref.~\cite{hdo25}. For shear-free collapse, $Y_{TF}$ is the difference between the electric part of the Weyl tensor and the pressure anisotropy, so that a vanishing complexity factor together with conformal flatness implies pressure isotropy. This is the setting of the radiating models of Herrera, Le Denmat, Santos and Wang \cite{hlsw04}, for which Maharaj and Govender \cite{mg05} solved the junction condition exactly. The structure scalars arise from the orthogonal splitting of the Riemann tensor \cite{bel,gomezlobo,h09}, and the term complexity itself has a statistical counterpart \cite{lopezruiz95} from which the idea of a reference system of vanishing complexity is borrowed.

The complexity factor has been applied to dynamical radiating models by Bogadi et al. In the shear-free, separable setting with an initially static progenitor, Bogadi and Govender \cite{bg22} showed that the dissipative term in $Y_{TF}$ is cancelled by the time-dependent part of the density gradient, so that the complexity factor reduces to a sum of pressure anisotropy and density inhomogeneity. Their static core obeys a linear equation of state with $\alpha=1/3$, so that the surface density is $4B$ with $B$ the bag constant, and they found that, as the star loses hydrostatic equilibrium, complexity increases and departs from zero as collapse proceeds. The evolution of complexity was found to track the Tolman--Oppenheimer--Volkoff forces in the collapsing fluid, a connection also explored in Ref.~\cite{bgm21}. Bogadi et al. \cite{bgm22} adopted the opposite stance and assumed that the dynamical complexity factor vanishes, in order to establish the restrictions this imposes on radiating models, noting that vanishing complexity is unlikely to persist in a collapsing, dissipating system. More recently, a perturbative treatment in which vanishing complexity closes the system of equations during dissipative collapse was given by Govender et al. \cite{gbgm24}. The role of shear has been addressed by Reddy and Govender \cite{rg26}, who examined its impact on the Weyl stresses and on the complexity factor in a collapsing core that dissipates heat to a Vaidya exterior, using a model in which the shear can be switched off, and demonstrated how shear alters the complexity and its components.

 Herrera \cite{h20} showed that energy-density inhomogeneity, dissipation and shear drive an initially isotropic configuration towards pressure anisotropy, so that pressure isotropy is not a stable condition and must be checked at every epoch. This is closely related to the instability of the shear-free condition itself \cite{hdo10}. Departures from the shear-free profile are driven by density inhomogeneity, heat flux and anisotropy, so that shear-free, isotropic models can be regarded as valid only for a limited epoch. Anisotropy is therefore the natural state of a collapsing, dissipating fluid, and, as we show below, the anisotropy of the models considered here is fixed by the static progenitor.

Another strand of the literature concerns the thermal behaviour of radiating stars. When the heat flux is determined by the Fourier (Eckart) law, signals propagate acausally and relaxational effects are ignored. Govender et al. \cite{gmm98} replaced this by a causal Maxwell--Cattaneo transport equation in a simple shear-free, geodesic model. Govender et al. \cite{gmm99} then studied relaxational effects for collapse starting from a static configuration and found that causal temperature profiles can differ markedly from the Eckart prediction. Maharaj and Govender \cite{mg05} showed that the same is true for collapse with vanishing Weyl stresses. Naidu et al. \cite{naidu06} and Govender, Reddy and Maharaj \cite{grm14} extended these results to shearing and anisotropic collapse, where the relaxation time of the shear stress plays a significant role. In all of these works the causal temperature at the core exceeds the non-causal one for a substantial part of the evolution, and depends on the particular static configuration chosen.

Realistic stellar matter obeys an equation of state (EoS), and the role of the EoS in radiating stars has attracted growing interest. Wagh et al. \cite{wagh01} imposed the linear relation $p=\alpha\rho$ on shear-free spacetimes with a linear temporal dependence. Sharma and Maharaj \cite{sm07} constructed static stellar models obeying a linear EoS, and Abebe et al. \cite{abebe14} used Lie analysis to obtain Euclidean stars with an EoS. Brassel et al. \cite{bgm17} analysed the collapse of radiating stars with several equations of state in a generalised Vaidya spacetime, and found that the final state is a locally naked central singularity. Govender et al. \cite{govender} considered a star which collapses from an initial static configuration with a linear EoS, $p_r=\alpha\rho-\eta$, and imposed the same EoS on the dynamical fluid. They reported that the EoS parameter alters the thermal profiles and the end state of collapse. A perturbative treatment of a linear EoS in an initially static Schwarzschild star was given by Govender et al. \cite{gmh19}, who found that the stability of the core depends on the EoS parameter. More recently, Naidoo et al. \cite{nmg23} analysed the junction condition for radiating stars, building on the transformation approaches of Thirukkanesh and Maharaj, Ivanov and others, and examined how supplementing the boundary condition with an EoS affects the evolution of the star. They showed that a linear EoS appears in the solutions of the boundary condition only under specific constraints, and that the EoS parameter, which also models exotic matter such as dark-energy stars with negative pressure, influences the stellar dynamics. In that analysis the EoS emerges from, or is compatible with, the solution of the boundary condition. This differs from imposing it a priori on the dynamical fluid, and the distinction is central to the present comment.

A physically motivated linear EoS of the form $p_r=\alpha\rho-\eta$ is the MIT bag model of quark matter \cite{chodos,farhi,witten}, in which massless, non-interacting quarks are confined by a vacuum pressure $B$ (the bag constant) and
\begin{equation}
p=\tfrac13\left(\rho-4B\right),
\end{equation}
which corresponds to $\alpha=1/3$ and $\eta=4B/3$. Imposing this relation on the initial static core, with $(p_r)_s$ vanishing at the boundary, gives $\eta_s=\alpha(\rho_s)_\Sigma$ and hence $(\rho_s)_\Sigma=4B$: the surface density of the static progenitor is fixed by the bag constant, as for strange quark stars. Static anisotropic models with a linear EoS were obtained by Sharma and Maharaj \cite{sm07}, and the bag EoS has been used in the setting of collapse with a null-fluid atmosphere. Harko and Cheng \cite{harko00} found exact solutions for collapsing strange quark matter in Vaidya geometry, and Ghosh and Dadhich \cite{ghosh03} studied the implications for cosmic censorship. In these models $B$ is a property of the QCD vacuum and hence a constant. It is therefore of interest to ask whether a bag-type relation, with fixed $\eta$, can be maintained by a radiating star throughout its dissipative evolution, or only at the initial static epoch where it determines $\eta_s$. We return to this question below.

Assuming a shear-free, separable metric for the interior given by
\begin{equation}
ds^2=-A_0^2(r)\,dt^2+f^2(t)\left[B_0^2(r)\,dr^2+r^2 d\Omega^2\right],
\end{equation}
where $(A_0(r),B_0(r))$ represents the initial static configuration, the Santos junction condition, continuity of the radial pressure and the heat flux at the
surface, $p_\Sigma=(qB)_\Sigma$, yields a single temporal equation for $f(t)$ \cite{santos}.
The work of Ref.~\cite{govender} prescribes a linear EoS
\begin{equation} \label{eta} p_r=\alpha\rho-\eta\end{equation} for
the initial static core to determine $A_0(r)$, and then studies the ensuing dynamics for dust
($\alpha=0$), radiation ($\alpha=1/3$), stiff ($\alpha=1$) and dark ($\alpha=-1/3$) fluids. In (\ref{eta}), $\eta > 0$ is a constant.
There the dynamical fluid is also required to obey $p_r=\alpha\rho-\eta$, and combination
with the junction condition is reported to give $\alpha=-1/3$. We show that this conclusion
holds for any interior within the separable ansatz $A=A_0(r)$, $B=f(t)B_0(r)$, identify a step in the derivation that is not justified, and discuss
its consequences for models that vary $\alpha$.

\section{Interior spacetime and energy--momentum tensor}
We adopt the framework of Ref.~\cite{govender}. The interior of a spherically symmetric,
shear-free radiating star is described by the line element
\begin{equation}
\begin{split}ds^2_-=&-A_0^2(r)\,dt^2\\ &+f^2(t)\left[B_0^2(r)\,dr^2+r^2\,d\Omega^2\right],\end{split} \label{int}
\end{equation}
where $d\Omega^2=d\theta^2+\sin^2\theta\,d\phi^2$, $(A_0,B_0)$ are the metric potentials of the initial static configuration and $f(t)$
describes the temporal evolution, with $f=1$ at the initial static epoch (in geometrized units). The exterior of the radiating star is described by the Vaidya spacetime \cite{vaidya},
\begin{equation}
\begin{split}ds^2_+=&-\left(1-\frac{2m(v)}{\mathcal R}\right)dv^2-2\,dv\,d\mathcal R\\ &+\mathcal R^2 d\Omega^2,\end{split} \label{ext}
\end{equation}
where $m(v)$ is the mass observed by an observer located at infinity and $v$ is retarded time coordinate.

For the interior, the matter content is described by an anisotropic fluid with a radial heat flux, with
energy--momentum tensor
\begin{equation}
\begin{split}T_{\alpha\beta}=&(\rho+p_t)u_\alpha u_\beta+p_t g_{\alpha\beta}+(p_r-p_t)\chi_\alpha\chi_\beta\\ &+q_\alpha u_\beta+q_\beta u_\alpha,\end{split} \label{emt}
\end{equation}
where $\rho$ is the energy density, $p_r$ and $p_t$ are the radial and tangential
pressures, and $q^\alpha$ is the heat-flow vector, taken along the radial direction by spherical symmetry. The fluid four-velocity is comoving,
\begin{equation}
u^\alpha=\frac{1}{A_0}\,\delta^\alpha_0, \label{u}
\end{equation}
$\chi^\alpha$ is a unit spacelike vector along the radial direction, and the vectors satisfy
$u_\alpha u^\alpha=-1$, $\chi_\alpha\chi^\alpha=1$ and $u_\alpha\chi^\alpha=u_\alpha q^\alpha=0$. The four-acceleration is given by
\begin{equation}
a_\alpha = \left(0, \frac{A_0'}{A_0}, 0, 0\right)
\end{equation} which is orthogonal to the four-velocity.

\section{Shear-free model}
For the line element~(\ref{int}) and energy--momentum tensor~(\ref{emt}) the Einstein field
equations $G_{\alpha\beta}=8\pi T_{\alpha\beta}$ can be cast as  \cite{govender}
\begin{align}
8\pi\rho &= \frac{1}{f^2}\left[\frac{1}{r^2}-\frac{1}{r^2B_0^2}+\frac{2B_0'}{rB_0^3}\right]
 +\frac{3\fd^2}{A_0^2f^2}, \label{e1}\\
8\pi p_r &= \frac{1}{f^2}\left[-\frac{1}{r^2}+\frac{1}{r^2B_0^2}+\frac{2A_0'}{rA_0B_0^2}\right] \nonumber\\
 &\quad -\frac{1}{A_0^2}\left[\frac{2\fdd}{f}+\frac{\fd^2}{f^2}\right], \label{e2}\\
8\pi p_t &= \frac{1}{f^2}\left[\frac{A_0''}{A_0B_0^2}+\frac{A_0'}{rA_0B_0^2}-\frac{B_0'}{rB_0^3}-\frac{A_0'B_0'}{A_0B_0^3}\right] \nonumber\\
 &\quad -\frac{1}{A_0^2}\left[\frac{2\fdd}{f}+\frac{\fd^2}{f^2}\right], \label{e3}\\
8\pi q^1 &= -\frac{2A_0'\fd}{A_0^2B_0^2f^3}, \label{e4}
\end{align}
where primes denote $\partial/\partial r$ and dots $\partial/\partial t$. These equations can be written more compactly as 
\begin{align}
\rho &= \frac{\rho_s}{f^2}+\frac{3\fd^2}{8\pi A_0^2f^2}, \label{rho}\\
p_r &= \frac{(p_r)_s}{f^2}-\frac{1}{8\pi A_0^2}\left[\frac{2\fdd}{f}+\frac{\fd^2}{f^2}\right], \label{pr}\\
p_t &= \frac{(p_t)_s}{f^2}-\frac{1}{8\pi A_0^2}\left[\frac{2\fdd}{f}+\frac{\fd^2}{f^2}\right], \label{pt}\\
q^1 &= -\frac{2A_0'\fd}{8\pi A_0^2B_0^2f^3}, \label{q}
\end{align}
where $\rho_s$, $(p_r)_s$ and $(p_t)_s$ are the energy density, radial pressure and
tangential pressure of the static configuration 
\begin{align}
8\pi\rho_s &= \frac{1}{r^2}-\frac{1}{r^2B_0^2}+\frac{2B_0'}{rB_0^3}, \label{s1}\\
8\pi(p_r)_s &= -\frac{1}{r^2}+\frac{1}{r^2B_0^2}+\frac{2A_0'}{rA_0B_0^2}, \label{s2}\\
8\pi(p_t)_s &= \frac{A_0''}{A_0B_0^2}+\frac{A_0'}{rA_0B_0^2}-\frac{B_0'}{rB_0^3}-\frac{A_0'B_0'}{A_0B_0^3}. \label{s3}
\end{align} regained when $f=1$, $\fd=\fdd=0$.
The static quantities depend solely on $A_0$ and $B_0$ only, and the time dependence of the
dynamical fluid is encoded entirely by $f(t)$ and its derivatives. The anisotropy
$\Delta=p_r-p_t=\Delta_s(r)/(8\pi f^2)$ therefore scales as $f^{-2}$ and depends on time only
through this prefactor. 

\section{Junction conditions}
The interior metric~(\ref{int}) is matched to the Vaidya exterior~(\ref{ext}) across a
timelike hypersurface $\Sigma$ ($r=r_\Sigma$) using the Darmois--Israel conditions first derived by Santos \cite{santos}. Continuity of the first fundamental form gives
\begin{equation}
\left(A_0\,dt\right)_\Sigma=d\tau,\qquad \mathcal R_\Sigma=\left(rB_0f\right)_\Sigma, \label{first}
\end{equation}
with $\tau$ the proper time on $\Sigma$ and $\mathcal R=rB_0f(t)$ the areal radius. The
Misner--Sharp mass \cite{ms} evaluated on $\Sigma$ then fixes the exterior mass,
\begin{equation}
m(v)=\frac{rf}{2}\left[1-\frac{1}{B_0^2}+\frac{r^2\fd^2}{A_0^2}\right]_\Sigma. \label{mass}
\end{equation}
Continuity of the second fundamental form leads to the condition that the momentum flux is
conserved across $\Sigma$, i.e.\ the radial pressure at the boundary is balanced by the heat flux,
\begin{equation}
p_\Sigma=(q^1B)_\Sigma, \label{santosc}
\end{equation}
where $B=fB_0$ and $q = (q^1q_1)^{1/2}$ is the magnitude of the heat flux. A nonvanishing radial pressure at the
surface is thus required whenever the star radiates. For the initial static configuration we require that the radial pressure vanishes at the boundary, $(p_r)_s|_\Sigma=0$ which is equivalent to \[(p_r)_s=\alpha\rho_s-\eta_s\]
 yielding 
$\eta_s=\alpha(\rho_s)_\Sigma$. Eqs.~(\ref{pr}) and~(\ref{q}) turn~(\ref{santosc}) into the
temporal evolution equation
\begin{equation}
2f\fdd+\fd^2-2n\fd=0,\qquad n\equiv\left(\frac{A_0'}{B_0}\right)_\Sigma. \label{junc}
\end{equation}
This single equation determines $f(t)$ for any choice of $A_0$ and $B_0$.

\section{Linear equation of state}
Suppose the static configuration obeys $(p_r)_s=\alpha\rho_s-\eta_s$ and the evolving fluid
obeys $p_r=\alpha\rho-\eta$, with $\eta$ constant, for all $r$ and $t$. Substituting
Eqs.~(\ref{rho}) and (\ref{pr}), the terms $\alpha\rho_s/f^2$ cancel, and after multiplication
by $8\pi A_0^2f^2$ we obtain
\begin{equation}
2f\fdd+(1+3\alpha)\fd^2=8\pi A_0^2(r)\left[\eta f^2(t)-\eta_s\right]. \label{eos}
\end{equation}
The left-hand side depends on $t$ alone. If $A_0(r)$ is not constant (and it cannot be, since
$q\propto A_0'$ vanishes otherwise), the right-hand side is $r$-independent only if
$\eta f^2(t)=\eta_s$ for all $t$. Since $f(t)$ is not constant during collapse, this requires
\begin{equation}
\eta=\eta_s=0. \label{eta0}
\end{equation}
This is not highlighted in \cite{govender}, where the $\eta$ terms are dropped in passing from the
dynamical equation of state to the equation for $\fd$. With (\ref{eta0}), Eq.~(\ref{eos}) becomes
\begin{equation}
2f\fdd+(1+3\alpha)\fd^2=0. \label{star}
\end{equation}

Let us now investigate a dynamical $\eta=\eta(t)$, with $\eta_s$ still a constant fixed by the static core. Eq.~(\ref{eos}) holds with $\eta$ replaced by $\eta(t)$. The left-hand side is
again a function of $t$ alone, so for non-constant $A_0(r)$ the bracket on the right must be
independent of $r$, which requires
\begin{equation}
\eta(t)\,f^2(t)=\eta_s. \label{etat}
\end{equation}
That is, $\eta(t)=\eta_s/f^2(t)$, and $\eta_s$ need not vanish. The right-hand side of
Eq.~(\ref{eos}) then vanishes identically, so Eq.~(\ref{star}) holds exactly as before. The
constant-$\eta$ result (\ref{eta0}) is the special case in which $\eta f^2=\eta_s$ can hold for
non-constant $f$ only if both sides vanish. Note that (\ref{etat}) is not an independent
freedom: $\eta(t)$ is fixed entirely by $f(t)$ and the static core, so letting $\eta$ vary in
time adds no new solutions to the junction problem.

\section{Result: $\alpha=-1/3$}
Subtracting (\ref{star}) from (\ref{junc}) gives $3\alpha\fd^2=-2n\fd$, i.e.\
$\fd(3\alpha\fd+2n)=0$. For $\fd\neq0$,
\begin{equation}
\fd=-\frac{2n}{3\alpha}, \label{fd}
\end{equation}
which is constant, so $\fdd=0$. Eq.~(\ref{junc}) then reduces to $\fd^2=2n\fd$, hence $\fd=2n$.
Equating with (\ref{fd}),
\begin{equation}
\alpha=-\tfrac13\qquad(n\neq0). \label{alpha}
\end{equation}

\begin{proposition}
For any separable shear-free radiating interior with non-constant $A_0(r)$ and arbitrary
$B_0(r)$, a linear equation of state $p_r=\alpha\rho-\eta$ holding for the whole evolution is
compatible with the Santos junction condition only for $\alpha=-1/3$, with $\eta=0$ if $\eta$
is constant and $\eta(t)=\eta_s/f^2(t)$ if $\eta$ is allowed to depend on time.
\end{proposition}
The excluded alternatives are trivial: $\alpha=0$ forces $n\fd=0$, $n=0$ means no radiation at
the boundary, and $\fd=0$ is the static case.

\textit{Remark (bag model).} For the MIT bag EoS, $\alpha=1/3$ and $\eta=4B/3$ with $B$ constant. The Proposition then excludes this relation for the dynamical fluid: $\alpha=1/3\neq-1/3$, and a constant $\eta$ would in any case have to vanish, which means $B=0$. The only freedom, $\eta(t)=\eta_s/f^2(t)$, would make the effective bag constant time dependent, in conflict with its interpretation as a vacuum pressure. The bag relation can therefore be imposed only on the initial static configuration, where it fixes $\eta_s=\tfrac43 B$ and the surface density $(\rho_s)_\Sigma=4B$.

\section{Consistency with the collapse solution}
The general solution of (\ref{junc}) is
\begin{equation}
t=\frac1n\left[\frac f2+\sqrt f+\ln\!\left(1-\sqrt f\right)\right], \label{tsol}
\end{equation}
with
\begin{equation}
\fd=-\frac{2n\left(1-\sqrt f\right)}{\sqrt f}, \label{fdsol}
\end{equation}
which is non-constant and negative for $n>0$ and $f<1$. The $\alpha=-1/3$ branch instead has
$\fdd=0$ and $\fd=+2n$, which for $n>0$ describes expansion at constant rate rather than
collapse. Thus the equation-of-state branch is not a member of the family
(\ref{tsol})--(\ref{fdsol}) used in \cite{govender} to compute density, pressure and temperature
profiles for $\alpha=1,\,1/3,\,0$ and $-1/3$.

The practical consequence is that a model which uses (\ref{tsol}) together with several values
of $\alpha$ cannot also claim the dynamical fluid obeys $p_r=\alpha\rho-\eta$. The only
consistent interpretation is that the equation of state holds on the initial static slice, where
it fixes $A_0(r)$ through $(p_r)_s=\alpha\rho_s-\eta_s$ and $\alpha$ is a free parameter. The
temperature profiles then depend on $\alpha$ through $A_0(r)$, and no restriction on $\alpha$
arises from the junction condition. Statements such as ``the equation of state is obeyed
throughout the collapse'' should therefore be dropped.

\section{Conclusion}
We have shown, for arbitrary separable potentials $A=A_0(r)$, $B=f(t)B_0(r)$, that a linear equation of state imposed on the dynamical fluid
of a shear-free radiating star is compatible with the Santos junction condition only for
$\alpha=-1/3$ (with $\eta=0$, or $\eta=\eta_s/f^2$ if $\eta$ varies in time), and that this branch
is non-collapsing. Studies of such models should impose the equation of state on the static core
only, or relax shear-freeness, the separability of the metric, or the matching assumptions, to
obtain dynamical equations of state that are genuinely compatible with dissipative collapse.

\textit{Role of separability.} The result relies on separability. For a general shear-free interior
$A=A(r,t)$, $B=B(r,t)$, the linear EoS throughout the interior is a single partial differential
equation relating $A$ and $B$, which no longer splits into an $r$-part and a $t$-part, so no ordinary
differential equation for $t$ alone follows. The Santos condition, by contrast, constrains only the
boundary values at $r=r_\Sigma$. The two are then compatible without fixing $\alpha$, since free functions remain.
The conclusion $\alpha=-1/3$ should therefore not be extended beyond the separable class without a
separate analysis.

\textit{Shearing case.} The distinction between a surface condition and an interior condition persists when shear is included. The Santos condition $p_r=q$, in its shearing form, is a statement about quantities evaluated on the comoving surface $r=r_\Sigma$: it constrains the interior solution only through its values and radial derivatives at $r_\Sigma$, and yields an ordinary differential equation governing the temporal behaviour of the model. A linear EoS imposed on the interior, by contrast, is a partial differential equation holding for all $0 \le r\le r_\Sigma$. The two requirements are therefore of different kinds, and in general, with shear, they can be satisfied together without restricting $\alpha$, since free functions of $r$ remain in the interior. Treating the junction condition as holding for all $r$ and $t$, and combining it with the equation of state as a single system of partial differential equations (as in Ref.~\cite{nmg23}), replaces the surface condition by the much {\em stronger requirement} $p_r=q$ throughout the star. This ties the heat flux to the density everywhere, and is difficult to reconcile with a regular centre, where $q\to0$ while $p_r=\alpha\rho-\eta$ is generally nonzero. Solutions obtained in this way are solutions of a more restrictive problem than the matching problem, and should not be read as showing that a radiating star with an interior EoS and a proper junction condition exists. A consistent treatment, with or without shear, should impose the EoS in the interior and the junction condition only at $r=r_\Sigma$. The junction condition can be used to determine the temporal behaviour of the free functions arising from the solution of the EoS constraint equation for the interior.

\section{Data Availability Statement}
This manuscript has no associated data
or the data will not be deposited. [There is no observational data related
to this article. The necessary calculations and graphic discussion can be
made available on request.]
\section{Code Availability Statement}
This manuscript has no associated
code/software. [Authors’ comment:Code/Software sharing not applicable
to this article as no code/software was generated or analysed during
the current study.]

\end{document}